\documentclass[prb,aps,amssymb,noshowpacs,twocolumn]{revtex4}
\usepackage{bbm}
\usepackage{mathrsfs}
\usepackage{amsfonts}
\usepackage{mathrsfs}
\usepackage{graphicx}

\begin{document}

\title{Exotic phase diagram of a topological quantum system}
\author{Xiao-Feng Shi, Yan Chen, and J. Q. You}
\affiliation{Department of Physics and State Key Laboratory of
Surface Physics, Fudan University, Shanghai 200433, China}
%


\date{\today}

\begin{abstract}
We study the quantum phase transitions (QPTs) in the Kitaev spin
model on a triangle-honeycomb lattice. In addition to the ordinary
topological QPTs between Abelian and non-Abelian phases, we find new
QPTs which can occur between two phases belonging to the same
topological class, namely, either two non-Abelian phases with the
same Chern number or two Abelian phases with the same Chern number.
Such QPTs result from the singular behaviors of the nonlocal
spin-spin correlation functions at the critical points.
\end{abstract}
 \maketitle

\section{Introduction}

A quantum phase transition (QPT) involves an abrupt change of the
ground state in a many-body system due to its quantum
fluctuations.\cite{Sachdev} Discovering and characterizing new QPTs
in a two-dimensional topological quantum system have recently
attracted considerable interest.\cite{HongYao,XYFeng,Kitaev06,xgwen}
Because the ground states in some topological quantum systems (e.g.,
the Kitaev spin models on honeycomb\cite{Kitaev06} and
triangle-honeycomb\cite{HongYao} lattices) are exactly solvable,
QPTs in these systems can be analytically investigated. In these
topological systems, the discovered QPTs include the transition
between a gapped Abelian phase and a gapless
phase,\cite{XYFeng,SYang} the transition between Abelian and
non-Abelian phases,\cite{CNash,HongYao,FABais,JQYou,SBChung} and the
transition between two non-Abelian phases with different Chern
numbers.\cite{Kitaev06} Also, an unconventional QPT between two
non-Abelian phases was found\cite{GKells} in the Kitaev spin model
on a triangle-honeycomb lattice by a fermionization method.
Nevertheless, to the best of our knowledge, the QPT between two
topological phases of the same Chern number (which belong to the
same topological class) has not yet been found.

Here we show that, in the Kitaev spin model on a triangle-honeycomb
lattice, a QPT can indeed happen between two gapped phases in the
same topological class, in addition to the ordinary topological QPT
between two phases of different Chern numbers. To demonstrate this,
we focus on two parameter regimes as typical examples: (i) When the
parameters vary across a critical curve separating two gapped phases
of the same Chern number $\nu=0$ or $\pm1$, a first-order QPT
occurs; (ii) when the parameters vary across a special critical
point where several critical curves meet, a continuous QPT can
occur. This is due to the exotic ground-state phase diagram which
has \textit{either} critical curves between two gapped phases (with
the same or different Chern numbers) \textit{or} critical points
where four different gapped phases (with Chern numbers $0,~0,~1$ and
$-1$) terminate. These results reveal that the Kitaev spin model on
a triangle-honeycomb lattice exhibits novel topological properties.
Moreover, we find that such QPTs result from the singular behaviors
of the nonlocal spin-spin correlations at the critical points.

The paper is organized as follows. In Sec.~II, we present the
solution for the ground state of the Hamiltonian in the uniform-flux
sector. Sections III and IV show two typical phase diagrams of the
ground-state wavefunctions and study various QPTs in the considered
Kitaev spin model. Finally, a brief conclusion is given in Sec.~V.

\section{Ground state in the uniform-flux sector}

\begin{figure}
\includegraphics
[width=3.1in,bbllx=71,bburx=565,bblly=332,bbury=772] {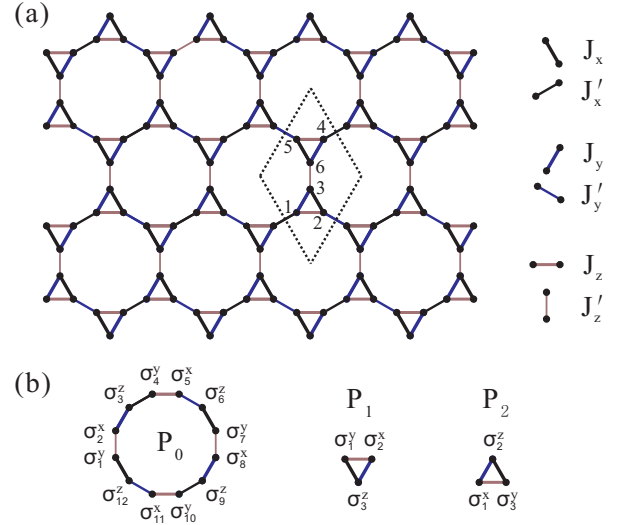}
 \caption{(Color online) (a) A schematic illustration of the Kitaev spin model on a
  triangle-honeycomb lattice. Each unit cell (dotted diamond) contains six spins at
  sites 1,2,$\cdots$,6. The six types of
bonds are labeled by $J_{\alpha}$ and $J_{\alpha}'$, where
$\alpha=x,y,z$. (b) Schematic diagrams of the three plaquette
operators $P_0,~P_1$, and $P_2$, which are defined in
Eq.~(\ref{plaquette}). }\label{Fisher}
\end{figure}

The Kitaev spin model on a triangle-honeycomb lattice is
schematically shown in Fig.~\ref{Fisher}(a) and the model
Hamiltonian is given by

\begin{eqnarray}
H&\!=\!&J_x\sum_{x\text{-link}}\sigma_i^x\sigma_j^x
+J_y\sum_{y\text{-link}}\sigma_i^y\sigma_j^y+J_z\sum_{z\text{-link}}\sigma_i^z\sigma_j^z\nonumber\\
&&+J_x'\sum_{x'\text{-link}}\sigma_i^x\sigma_j^x
+J_y'\sum_{y'\text{-link}}\sigma_i^y\sigma_j^y+J_z'\sum_{z'\text{-link}}\sigma_i^z\sigma_j^z,\nonumber\\\label{spinH}
\end{eqnarray}
where $\sigma_i^{\alpha}$, with $\alpha=x,y$ and $z$, are the three
Pauli operators at site $i$. Among the six sums in
Eq.~(\ref{spinH}), four involve interactions within each unit cell
(see Fig.~\ref{Fisher}): (i) The $x$-link couples either spins 2 and
3 or spins 5 and 6, (ii) the $y$-link couples either spins 1 and 3
or spins 4 and 6, (iii) the $z$-link couples either spins 1 and 2 or
spins 4 and 5, and (iv) the $z'$-link couples spins 3 and 6.
Nearest-neighbor unit cells are coupled by the $x'$- and $y'$-links.
As found in Ref. \onlinecite{HongYao}, the ground state of
Hamiltonian (\ref{spinH}) is at least 8 (6)-fold degenerate for the
Abelian (non-Abelian) phase.

For the three types of hermitian plaquette operators defined by [see
Fig.~\ref{Fisher}(b)]
\begin{eqnarray}
P_0&\equiv&
\sigma_1^{y}\sigma_2^{x}\sigma_3^{z}\sigma_4^{y}\sigma_5^{x}\sigma_6^{z}\sigma_7^{y}\sigma_8^{x}\sigma_9^{z}
\sigma_{10}^{y}\sigma_{11}^{x}\sigma_{12}^{z} ,\nonumber\\
P_1&\equiv&\sigma_1^{y}\sigma_2^{x}\sigma_3^{z},~~ P_2\equiv
\sigma_1^{x}\sigma_2^{z} \sigma_3^{y},\label{plaquette}
\end{eqnarray}
each has eigenvalues $\pm1$. These plaquette operators commute with
not only each other but also the Hamiltonian (\ref{spinH}). As
verified in Ref. \onlinecite{HongYao}. the ground state of
Hamiltonian (\ref{spinH}) is \textit{either} in the sector of the
Hilbert space in which all plaquette operators in
Eq.~(\ref{plaquette}) have eigenvalue one \textit{or} in the sector
where the time-reversal transformation of each plaquette operator in
Eq.~(\ref{plaquette}) has eigenvalue one. The former is denoted as
the uniform-flux sector.\cite{footnote1}

\begin{figure}
\includegraphics
[width=3.30in,bbllx=0,bblly=333,bburx=578,bbury=603]
{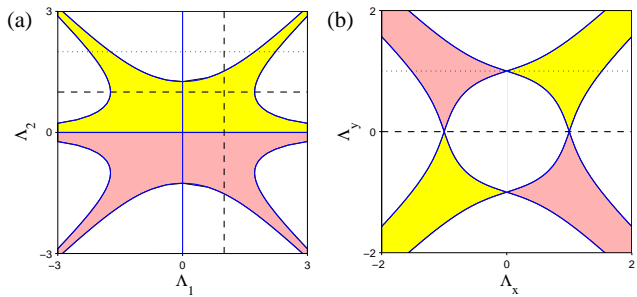}
 \caption{(Color online) The ground-state phase diagram of the
 Kitaev spin model on a triangle-honeycomb lattice. (a) $\Lambda_1=J_x'/J_x$, and $\Lambda_2=J_z/J_x$, when
 $J_x=J_y,~J_x'=J_y'$, and $J_zJ_x'=J_z'J_x$. (b) $\Lambda_x=J_x/J_z$, and $\Lambda_y=J_y/J_z$, when
 $J_x=J_x',~J_y=J_y'$, and $J_z =J_z'$. The solid curves or lines correspond to gapless phases. The yellow (red) regions in (a) and (b) correspond to gapped phases with Chern number $\nu=1$ ($-1$).
 The white regions are gapped
 phases with Chern number $\nu=0$.}\label{phasediagram}
\end{figure}

By using the Jordan-Wigner transformation, Hamiltonian (\ref{spinH})
can be converted to a form represented by Majorana
fermions.\cite{HongYao} Performing Fourier transform on the
Hamiltonian in the uniform-flux sector, we obtain
\begin{eqnarray}
H_u & =& \sum_{\mathbf{k}\in
\mathrm{BZ},j=1}^{j=3}\varepsilon_{\mathbf{k}}^{(j)}\left[2A_{\mathbf{k}}^{(j)\dag}A_{\mathbf{k}}^{(j)}-1\right],\label{Hfermion}
\end{eqnarray}
where $\mathrm{BZ}$ denotes the first Brillouin zone. In
Eq.~(\ref{Hfermion}), $
A_{\mathbf{k}}^{(i)\dag}=\sum_{s=1}^{6}c_{\mathbf{k}}^{(s)}w_{s} $
are fermionic operators, where $c_{\mathbf{k}}^{(s)}$ is the Fourier
transform of the Majorana fermionic operator at site $s$
($s=1,2,\cdots,6$) in a unit cell, and $w_{s} $ is a function of
variable $\varepsilon_{\mathbf{k}}^{(j)}$ (see Appendix A). One can
prove that Hamiltonian (\ref{Hfermion}) breaks time reversal
symmetry from the property of Majorana fermions
$c_{\mathbf{k}}^{(s)}$. This is in sharp contrast to the Kitaev
model on a honeycomb lattice in which the time reversal symmetry is
preserved. This is due to the difference between the bipartite
nature of the honeycomb lattice and the non-bipartite nature of the
triangle-honeycomb lattice. Hamiltonian (\ref{Hfermion}) has six
energy bands: $
\varepsilon_{\mathbf{k}}^{(1)}=-\varepsilon_{\mathbf{k}}^{(6)}=-\varepsilon_{\mathbf{k}},~\varepsilon_{\mathbf{k}}^{(2)}=
-\varepsilon_{\mathbf{k}}^{(5)}=-\varepsilon_{\mathbf{k}}^+$, and $
~\varepsilon_{\mathbf{k}}^{(3)}=-\varepsilon_{\mathbf{k}}^{(4)}=-\varepsilon_{\mathbf{k}}^-
$, with
$\varepsilon_{\mathbf{k}}\geq\varepsilon_{\mathbf{k}}^{+}\geq\varepsilon_{\mathbf{k}}^{-}\geq0$.
In each unit cell, six Majorana fermions are defined, but the number
of the corresponding fermions are three. Thus, the lowest three
bands should be filled for the ground state:
$|g\rangle=\prod_{\mathbf{k}}\prod_{j=1}^3\sqrt{2}A_{\mathbf{k}}^{(j)\dag}|0\rangle$,
with $\sqrt{2}$ the normalization factor which results from the fact
that the fermion $A_{\mathbf{k}}^{(j)}$ is constructed from Majorana
fermions in $\mathbf{k}$ space, each of whom can only take $1/2$ as
its occupation number. The ground-state energy per site is given by
$ E_g
=\frac{1}{6N}\sum_{\mathbf{k}}\left[\varepsilon_{\mathbf{k}}^{(1)}+\varepsilon_{\mathbf{k}}^{(2)}+\varepsilon_{\mathbf{k}}^{(3)}\right]$,
where $N$ is the number of unit cells. The energy-band gap, i.e.,
the minimal energy to excite a fermion from the ground state is
$\Delta=\mathscr{M}[\varepsilon_{\mathbf{k}}^{(4)}-\varepsilon_{\mathbf{k}}^{(3)}]$,
where $\mathscr{M}[\cdots]$ denotes the minimal value of a function
with variable $\mathbf{k}$.

Below we show the ground-state phase diagrams in two different
parameter regimes. The first diagram contains critical curves
separating two phases of either the same or different Chern numbers.
This unveils that QPTs can also occur in the same topological class.
The second diagram contains critical points where four different
phases with Chern numbers $0$, $0$, $1$ and $-1$ terminate.

\section{Case A: $J_x=J_y$, $J_x'=J_y'$, and
$J_zJ_x'=J_z'J_x$}

We first choose parameters $\Lambda_1\equiv J_x'/J_x$, and
$\Lambda_2\equiv J_z/J_x$ to study the ground-state property of the
system. In particular, the parameters with $\Lambda_2=1$ and
$\Lambda_1>0$ correspond to the case studied in Ref.
\onlinecite{HongYao} where a topological QPT between an Abelian
phase
 and a non-Abelian phase was found at point $\Lambda_1=\sqrt3$. There is
also a perturbative study\cite{SDusuel} of this model with
parameters either $\Lambda_1 \ll1$ or $\Lambda_1\gg1$ when
$\Lambda_2=1$.

In order to find the ground-state phase diagram of the Kitaev spin
model, we should first distinguish the gapless-phase regions from
the gapped-phase regions in the phase diagram. Then, we calculate
the Chern number\cite{XLQi06,TKNN} as a topological index to
characterize each gapped-phase region. We note that the six energy
bands satisfy the relation
$\varepsilon_{\mathbf{k}}^{(j)}=-\varepsilon_{\mathbf{k}}^{(7-j )}$,
where $j=1,2,3$, which implies that the closing of the energy-band
gap corresponds to either
$\mathscr{M}[-\varepsilon_{\mathbf{k}}^{(3)}]=0$ or
$\mathscr{M}[\varepsilon_{\mathbf{k}}^{(4)}]=0$. Combining this
condition with the relation
$\varepsilon_{\mathbf{k}}^{(j)}=-\varepsilon_{\mathbf{k}}^{(7-j )}$,
we can derive that if one has
$\mathscr{M}[-\prod_{j=1}^6\varepsilon_{\mathbf{k}}^{(j)}]=0$, the
energy-band gap is zero, i.e., $\Delta=0$. This gives rise to
\begin{eqnarray}
\Lambda_1 &=&0;~\Lambda_2 =0;~ \Lambda_1^2=\Lambda_2^2
\pm\frac{2}{|\Lambda_2|} ,\label{eqphase1}
\end{eqnarray}
where $+$ ($-$) applies when $|\Lambda_1|>|\Lambda_2|$
($|\Lambda_1|<|\Lambda_2|$). 
In Fig. \ref{phasediagram}(a), each gapless phase determined by
Eq.~(\ref{eqphase1}) is schematically shown by a solid curve or
line. One can see that the $\Lambda_1\Lambda_2$ plane is divided
into 12 distinct regions by these solid curves or lines. Each of
these 12 regions is a gapped topological phase and can be
characterized by a Chern number.

We can define the Chern number by using the Berry's phase for a
gapped ground state.\cite{TKNN,XLQi06} Among the six bands of
Hamiltonian (\ref{Hfermion}), the lower three bands with states $|
j,\mathbf{k}\rangle=\sqrt2A_{\mathbf{k}}^{(
j)\dag}\left|0\right\rangle$, where $j=1,2,3$, are occupied. The
Berry's phase gauge field in momentum space is
\begin{equation}
f_{\alpha}(\mathbf{k})=-i\sum_{ j=1}^{3}\left\langle
j,\mathbf{k}\left|\frac{\partial}{\partial k_{\alpha}}\right|
j,\mathbf{k}\right\rangle,
\end{equation}
where $\alpha=x,y$. This gives
 \begin{eqnarray}
 \nu &=&\frac{1}{\pi}\mathrm{Im}\sum_{s=1}^6\sum_{
i=1}^3\int_{\mathrm{BZ}}d^2k\left[ \frac{\partial
w_{s}^{\ast}(\varepsilon_{\mathbf{k}}^{( i)} )}{\partial
k_x}\frac{\partial w_{s}(\varepsilon_{\mathbf{k}}^{( i)} )}{\partial
k_y}\right],\label{Chernnumber}
\end{eqnarray}
where $\mathrm{Im}$ denotes the imaginary part of a complex
variable. Numerical results show that each yellow (red) region in
Fig.~\ref{phasediagram}(a) corresponds to a gapped non-Abelian
phases with Chern number $\nu=1 ~(-1)$. The other 8 white regions
correspond to the Abelian phases with $\nu=0$. A remarkable feature
 is that a critical curve or line (i.e. gapless phase) can separate two
gapped phases with Chern numbers $\nu=$ (i) 0 and $\pm1$, (ii) 0 and
0, or (iii) $1(-1)$ and $\pm1$. This reveals that a QPT can occur
between two gapped phases belonging to the same topological class.

\begin{figure}
\includegraphics[width=3.3in,bbllx=72,bblly=204,bburx=525,bbury=628]
{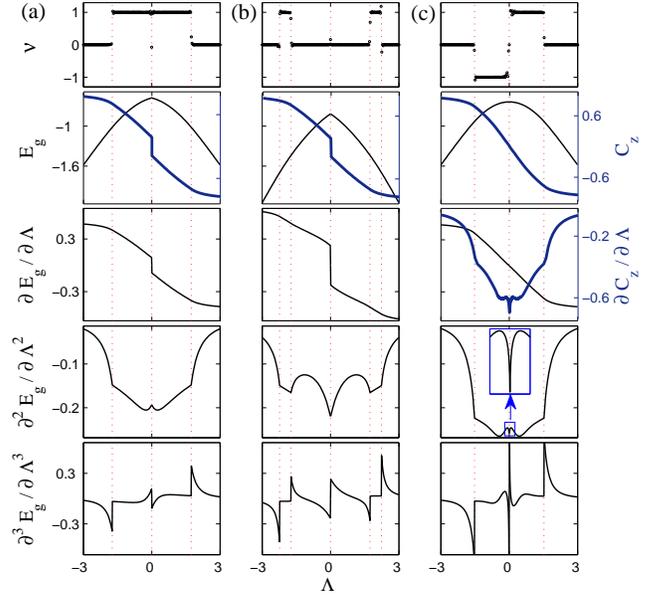}
 \caption{(Color online) Chern number $\nu$, the
 ground-state energy $E_g$ and its derivatives, and the
spin-spin correlation function $C_z$ and its derivative with respect
to the driving parameter. (a) $\Lambda=\Lambda_1$ and $\Lambda_2=1$;
(b) $\Lambda=\Lambda_1$ and $\Lambda_2=2$; (c) $\Lambda=\Lambda_2$
and $\Lambda_1=1$. The parameters in (a), (b) and (c) correspond to
the horizontal dashed, horizontal dotted and
 vertical dashed lines in Fig.~\ref{phasediagram}(a).}\label{ground01}
\end{figure}

Below we explicitly display the QPTs (see Fig. \ref{ground01}).
As in Ref. \onlinecite{Uzunov}, here we classify the QPTs as two
types: The \textit{first-order} QPT where the first derivative of
the ground-state energy $E_g$ with respect to the driving parameter
is discontinuous at the transition point, and the
\textit{continuous} QPT where a higher-order derivative of $E_g$ is
discontinuous and the derivative(s) with order(s) lower is (are)
continuous. Figure \ref{ground01}(a) [\ref{ground01}(b)] shows that
there is a first-order QPT between two phases of the \textit{same}
Chern number $\nu=1~(0)$ at $(\Lambda_1,~\Lambda_2)=(0,~1)$
[$(0,~2)]$, when one parameter varies along the horizontal dashed
(dotted) line in Fig.~\ref{phasediagram}(a). Using the perturbation
method in \onlinecite{Kitaev06}, the effective Hamiltonian at
$\Lambda_1\sim0$ can be obtained, up to third order, as (see
Appendix B)
\begin{eqnarray}
H_{\mathrm{eff}}\!&\!=\!&\!H_{0}^{(3)}+\frac{6\Lambda_1^3}{\Lambda_2}
\sum_{n}\left[P_1(n) \sigma_{i}^x\sigma_{j}^y\sigma_{k}^z+P_2(n)
\sigma_{i'}^x\sigma_{j'}^y\sigma_{k'}^z\right],\nonumber\\\label{effective}
\end{eqnarray}
where $n$ denotes the $n$\textit{th} unit cell, $H_{0}^{(3)}$ is a
term containing none of $P_0$, $P_1$ or $P_2$, and the subscripts
$i,j$, and $k$ ($i',j'$, and $k'$) denote the sites linked to
plaquette $P_1$ ($P_2$) with $x$-, $y$- and $z$-link, respectively.
For the QPT between two phases of the same Chern number, the
parameter $\Lambda_1$ changes its sign at the transition point
$\Lambda_1=0$. This gives rise to different $H_{\mathrm{eff}}$'s at
the two sides of the transition point $\Lambda_1=0$. Also, the
difference between the two phases of the same Chern number but with
positive and negative $\Lambda_1$'s can be seen from their
wavefunctions (see Appendix A).


Using Feynmann theorem,\cite{JQYou} it can be derived that
\begin{equation}
\frac{\partial E_g}{\partial
\Lambda_1}=\frac{1}{6}(C_x+C_y+\Lambda_2C_z),
\end{equation}
where the spin-spin correlation functions are defined by $C_{\alpha
}=\langle g|\sigma_i^{\alpha}\sigma_j^{\alpha}|g
\rangle_{\alpha'\text{-link}}$, with $\alpha=x,y$ and $z$ for
$\alpha'=x',y'$ and $z'$. It is clear that the discontinuity of
$\partial E_g/\partial \Lambda_1$ at a QPT point results from the
discontinuity of $C_{\alpha}$ there. Here we find that, at the QPT
point, the \textit{nonlocal} spin-spin correlation function on the
$x'$-, $y'$-, or $z'$-link changes its sign, displaying
discontinuity there. As shown in Fig. \ref{ground01}(a) [Fig.
\ref{ground01}(b)], $C_z$ indeed has a jump at the QPT point
$(\Lambda_1,~\Lambda_2)=(0,~1)$ [$(0,~2)]$ when $\Lambda_1$ changes
from positive to negative. It should be noted that only part of the
spin-spin couplings in the Hamiltonian (\ref{spinH}) change their
sign at these transition points, while the other spin-spin couplings
are kept unchanged.

In addition to the first-order QPTs, continuous QPTs can also occur
in the Kitaev spin model on a triangle-honeycomb lattice. As shown
in Fig. \ref{ground01}~(a), $\partial E_g/\partial \Lambda_1$ and
$\partial^2 E_g/\partial \Lambda_1^2$ are continuous while
$\partial^3 E_g/\partial \Lambda_1^3$ becomes discontinuous at
$(\Lambda_1,~\Lambda_2)=(\pm\sqrt3,~1)$. This corresponds to a
topological QPT between Abelian and non-Abelian phases.
Figure \ref{ground01}(b) shows that such a continuous QPT can also
happen at $(\Lambda_1,~\Lambda_2)=(\pm\sqrt3,~ 2)$ or
$(\pm\sqrt5,~2)$.
Similar to the first-order QPTs, the continuous QPTs result from the
discontinuity of the second derivative of spin-spin correlation
functions at the critical point.

Figure \ref{ground01}(c) displays three continuous QPTs. 
Interestingly, the QPT at $(\Lambda_1,~\Lambda_2)=(1,~ 0)$, where
$\partial^2 E_g/\partial \Lambda_2^2$ diverges, occurs between two
non-Abelian phases with Chern numbers $1$ and $-1$, respectively.
This is in sharp contrast to the other two QPTs occurring at
$\Lambda_1=1$ and $\Lambda_2\approx\pm1.5$, where $\partial^3
E_g/\partial \Lambda_2^3$ is discontinuous and each transition is
between Abelian and non-Abelian phases, instead of two non-Abelian
phases. Also, it can be derived that
\begin{equation}
\frac{\partial E_g}{\partial
\Lambda_2}=\frac{1}{6}(B_z+\Lambda_1C_z),
\end{equation}
where $B_z=\langle g|\sigma_i^z\sigma_j^z|g
\rangle_{z\text{-link}}$. This unveils that the nonanalyticity of
$E_g$ results from the nonanalyticity of either $B_z$ or $C_z$.
Indeed, Fig. \ref{ground01}(c) shows that $\partial C_z/\partial
\Lambda_2$ is divergent at $(\Lambda_1,~\Lambda_2)=(1,~ 0)$. Such a
QPT between two phases of Chern numbers $\nu=\pm1$ can also occur
when changing the sign of the magnetic-field-related parameter
$\kappa$ in the Kitaev spin model on a honeycomb
lattice.\cite{Kitaev06}

\begin{figure}
\includegraphics [width=3.4in,bbllx=43,bblly=196,bburx=567,bbury=631]
{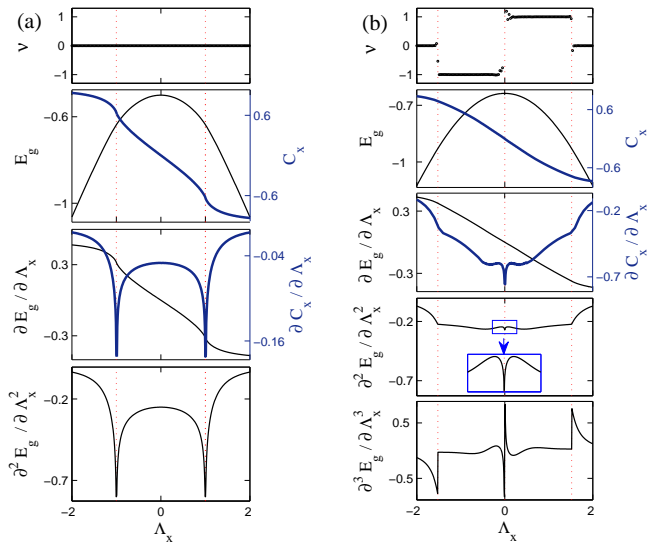}
 \caption{(Color online)
 Chern number $\nu$, the
 ground-state energy $E_g$, the spin-spin correlation function $C_x$, and the derivatives of both
 $E_g$ and $C_x$ with respect to the driving
 parameter $\Lambda_x$. The parameters in (a) and (b) correspond to the
 dashed and
 dotted lines in Fig.~\ref{phasediagram}(b).}\label{ground02}
\end{figure}

\section{Case B: $J_{\alpha}=J_{\alpha}'$, where
$\alpha=x$, $y$, and $z$}

Here we use parameters $\Lambda_x\equiv J_x/J_z$, and
$\Lambda_y\equiv J_y/J_z$ to characterize the phase diagram. The
gapless-phase curves determined by
$\mathscr{M}[-\prod_{j=1}^{6}\varepsilon_{\mathbf{k}}^{(j)}]=0$ are
\begin{eqnarray}
|\Lambda_y|&=&\sqrt[3]{T_2^++T_1^+}
+\sqrt[3]{T_1^+-T_2^+},~|\Lambda_y|>|\Lambda_x|; \nonumber\\
|\Lambda_y|&=&\sqrt[3]{T_2^--T_1^-}
-\sqrt[3]{T_2^-+T_1^-},~|\Lambda_x|>|\Lambda_y|;\\
|\Lambda_y|&=&\sqrt[3]{T_2^-+T_1^-}
-\sqrt[3]{T_2^--T_1^-},~1\geq|\Lambda_x|,|\Lambda_y|.\nonumber
\label{phase2}
\end{eqnarray}
Here $T_1^{\pm}=(1\pm|\Lambda_x|^3)/2$, and $T_2^{\pm}= \sqrt{
(T_1^{\pm})^2\mp(|\Lambda_x|/3)^3}$. The gapless phase determined by
each condition in Eq.~(\ref{phase2}) is shown by a solid curve in
Fig \ref{phasediagram}(b). These gapless-phase curves divide the
$\Lambda_x\Lambda_y$ plane into 9 regions: Two phases of Chern
number $\nu=1~(-1)$, denoted by yellow (red) regions, and five
gapped phases of Chern number $\nu=0$, denoted by white regions. In
this phase diagram, four gapped phases with $\nu=0,~0,~1$ and $-1$
all terminate at a gapless-phase point $(\Lambda_x,~\Lambda_y)=(0,
\pm1)$ or $(\pm1, 0)$. Such a point is analogous to the eutectic
point in crystallography. Also, a similar critical point exists in
the phase diagram of the Haldane model.\cite{Haldane}

Below we focus on the QPTs at the points where different gapped
phases terminate. When $\Lambda_x$ varies along the dashed line in
Fig. \ref{phasediagram}(b), a continuous QPT occurs at point
$(\Lambda_x,~\Lambda_y)=(\pm1, 0)$, where $\partial^2 E_g/\partial
\Lambda_x^2$ diverges [see Fig. \ref{ground02}(a)]. Similar to the
QPT between two non-Abelian phases belonging to the same topological
class, this QPT involves two Abelian phases with the same Chern
number $\nu=0$. When $\Lambda_x$ varies along the dotted line in
Fig. \ref{phasediagram}(b), in addition to the continuous QPTs
between Abelian and non-Abelian phases (which occur at
$\Lambda_x\approx\pm1.5$ and $\Lambda_y=1$, where $\partial^3
E_g/\partial \Lambda_x^3$ are discontinuous), a continuous QPT
between two non-Abelian phases happens at point
$(\Lambda_x,~\Lambda_y)=(0, 1)$, where $\partial^2 E_g/\partial
\Lambda_x^2$ diverges. In contrast to the QPT between two
non-Abelian phases belonging to the same topological class, this
transition involves two non-Abelian phases with $\nu=\pm1$.
Analogous to the first-order QPTs occurring in the same topological
class (Fig. \ref{ground01}), the continuous QPTs here are also due
to the singularity of the nonlocal correlation functions at the
critical points (see, e.g., the thick solid curves in Fig.
\ref{ground02}).

\section{Conclusion}

We have studied QPTs in the Kitaev spin model on a
triangle-honeycomb lattice and revealed the exotic ground-state
phase diagram of this model. In addition to the ordinary topological
QPTs between Abelian and non-Abelian phases, we find new QPTs that
occur between two phases belonging to the same topological class.
Moreover, we show that such QPTs are due to the singular behaviors
of the nonlocal spin-spin correlation functions at the critical
points.

\begin{acknowledgments}
This work was supported by the National Basic Research Program of
China under Grant No. 2009CB929300, and the National Natural Science
Foundation of China under Grant No. 10625416.
\end{acknowledgments}

\appendix
\section{}
\subsection{Derivation of the ground state in the uniform-flux sector}

\begin{figure}
\includegraphics[width=3.0in]
{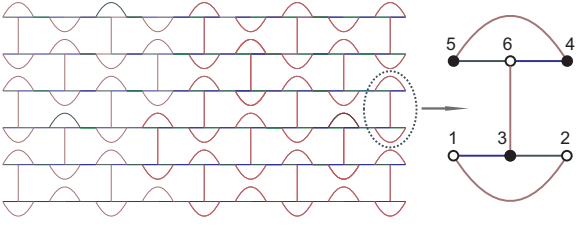} \caption{(Color online) The triangle-honeycomb lattice in
Fig.~\ref{Fisher}(a) is deformed to a topologically equivalent
lattice, so as to label each site by row and column indices. In each
unit cell, the three sites denoted by $3,~4$, and $5$ have row and
column indices $m$ and $n$ with $m+n$ equal to an even~integer, and
the other three sites denoted by $1,~2$, and $6$ have row and column
indices with $m+n$ equal to an odd~integer.}\label{Fisher2}
\end{figure}

 The Kitaev spin model
on a triangle-honeycomb lattice is schematically shown in
Fig.~\ref{Fisher2} and the model Hamiltonian is given by
\begin{eqnarray}
H&=&J_x\sum_{x\text{-link}}\sigma_i^x\sigma_j^x
+J_y\sum_{y\text{-link}}\sigma_i^y\sigma_j^y+J_z\sum_{z\text{-link}}\sigma_i^z\sigma_j^z\nonumber\\
&&+J_x'\sum_{x'\text{-link}}\sigma_i^x\sigma_j^x
+J_y'\sum_{y'\text{-link}}\sigma_i^y\sigma_j^y+J_z'\sum_{z'\text{-link}}\sigma_i^z\sigma_j^z.\nonumber\\\label{SspinH}
\end{eqnarray}
To use the Jordan-Wigner transformation, we need to label the sites
by row and column indices. We deform the honeycomb lattice [see
Fig~1(a)] into the topologically equivalent one in
Fig.~\ref{Fisher2}. Now each site can be labeled by $(m,n)$, where
$m$ ($=1,2,\cdots,M$) and $n$ ($=1,2,\cdots,N$) are the row and
column indices, respectively.

By performing the Jordan-Wigner transformation\cite{ELieb}
\begin{eqnarray}
\sigma_{m,n}^{+}&=&2a_{m,n}^{\dag}\prod_{n^{\prime}=1}^{N}\prod_{m^{\prime}<m}
\sigma_{m^{\prime},n'}^{z}
\prod_{n''<n}\sigma_{m,n''}^{z}\label{Jordan}
\end{eqnarray}
on each spin, and using the Majorana fermions
\begin{eqnarray}
c_{m,n}^{(s)}\equiv i(a_{m,n}^{\dag}-a_{m,n}),~d_{m,n}^{(s)}\equiv
a_{m,n}^{\dag}+a_{m,n}\label{Majorana1}
\end{eqnarray}
 for sites $s=3,4,5$, and
\begin{eqnarray}
c_{m,n}^{(s)}\equiv a_{m,n}^{\dag}+a_{m,n} ,~d_{m,n}^{(s)}\equiv
i(a_{m,n}^{\dag}-a_{m,n})\label{Majorana2}
\end{eqnarray}
 for sites $s=1,2,6$ in each unit cell [see
Fig.~\ref{Fisher2}], the Hamiltonian (\ref{SspinH}) is converted to
\begin{eqnarray}
H&\!=\!&iJ_x\left[\sum_{x\text{-link}(\bigtriangleup)}c_{m,n}^{(3)}c_{m,n+1}^{(2)}+\sum_{x\text{-link}(\bigtriangledown)}c_{m,n}^{(5)}c_{m,n+1}^{(6)}\right]
\nonumber\\&&-iJ_y\left[\sum_{y\text{-link}(\bigtriangleup)}c_{m,n}^{(1)}c_{m,n+1}^{(3)}+\sum_{y\text{-link}(\bigtriangledown)}c_{m,n}^{(6)}c_{m,n+1}^{(4)}\right]
\nonumber\\
&&+iJ_x'\sum_{x'\text{-link},n\neq
N}c_{m,n}^{(4)}c_{m,n+1}^{(1)}\nonumber\\
&&-iJ_y'\sum_{y'\text{-link},n\neq N}c_{m,n}^{(2)}c_{m,n+1}^{(5)}\nonumber\\&&+iJ_x'\sum_{m=1}^{M/2}\Phi_{x,2m}c_{2m,N}^{(4)}c_{2m,1}^{(1)}\nonumber\\
&&-iJ_y'\sum_{m=1}^{M/2}\Phi_{x,2m-1}c_{2m-1,N}^{(2)}c_{2m-1,1}^{(5)}\nonumber\\
&&+iJ_z
\sum_{\text{z-link}(\bigtriangleup)}id_{m,n}^{(1)}d_{m,n+2}^{(2)}c_{m,n}^{(1)}c_{m,n+2}^{(2)}\nonumber\\
&&+iJ_z\sum_{\text{z-link}(\bigtriangledown)}id_{m,n+2}^{(4)}d_{m,n}^{(5)}c_{m,n+2}^{(4)}c_{m,n}^{(5)}
\nonumber\\&&+iJ_z'\sum_{z'\text{-link}}id_{m,n}^{(3)}d_{m+1,n}^{(6)}c_{m,n}^{(3)}c_{m+1,n}^{(6)},\label{HMajorana}
\end{eqnarray}
where $\Phi_{x,m}=\prod_{n=1}^N\sigma_{m,n}^z$, $m=1,2,\cdots$, or
$M$, is the global flux operator calculated along a given contour
encircling the system in the horizontal direction,
$\alpha\text{-link}(\bigtriangleup)$ denotes the $\alpha\text{-link}
$ in an up-pointing triangle, and
$\alpha\text{-link}(\bigtriangledown)$ the $\alpha\text{-link} $ in
a down-pointing triangle.

The global flux operator calculated along a given contour encircling
the system in the vertical direction, which does not appear in the
Hamiltonian (\ref{HMajorana}), is given by
\begin{eqnarray}
\Phi_{y,n}&=&\prod_{m=0}^{M/2-1}\sigma_{1+2m,n}^y\sigma_{2+2m,n}^x\sigma_{2+2m,n+1}^z\sigma_{2+2m,n+2}^z\nonumber\\
&&\otimes\sigma_{2+2m,n+3}^x\sigma_{3+2m,n+3}^y\sigma_{3+2m,n+2}^z\sigma_{3+2m,n+1}^z\nonumber\\
&=&(-1)^{\frac{M}{2}}\prod_{m=0}^{M/2-1}d_{1+2m,n}^{(3)}d_{2+2m,n}^{(6)}d_{2+2m,n+3}^{(3)}\nonumber\\
&&\otimes d_{3+2m,n+3}^{(6)},\label{phaseY}
\end{eqnarray}
where $n=1+6r$ with $r=0,1,\cdots$, or $N/6-1$. These two global
flux operators $\Phi_{x,m}$ and $\Phi_{y,n}$ commute with
Hamiltonian (\ref{SspinH}). This leads to a topological degeneracy
for the ground state because $\Phi_{x,m}$ and $\Phi_{y,n}$ can have
eigenvalues either $1$ or $-1$. For the Abelian phase, this
degeneracy is 4-fold  because $(\Phi_{x,m},\Phi_{y,n})$ can be
$(1,1)$, $(1,-1)$, $(-1,1)$, and $(-1,-1)$, while the degeneracy is
3-fold for the non-Abelian phase since
$(\Phi_{x,m},\Phi_{y,n})=(-1,-1)$ is not allowed.\cite{HongYao} In
addition to the global flux operators, the three types of local
plaquette operators $P_0,~P_1$, and $P_2$ also commute with
Hamiltonian (\ref{SspinH}). Using the Jordan-Wigner transform, they
can be written as
\begin{eqnarray}
P_{0}(m,n)&=&d_{m,n}^{(3)}d_{m+1,n}^{(6)}
d_{m,n+6}^{(3)}d_{m+1,n+6}^{(6)} \nonumber\\&& \otimes P_{1}(m,n+2)P_{2}(m+1,n+2),\nonumber\\
P_{1}(m',n')&=&
id_{m',n'+2}^{(4)}d_{m',n'}^{(5)},\nonumber\\
P_{2}(m'',n'')&=&id_{m'',n''}^{(1)}d_{m'',n''+2}^{(2)},\label{plaquetteA7}
\end{eqnarray}
where $(m,n)$, $(m',n')$, and $(m'',n'')$ correspond, respectively,
to the sites denoted by $3$, $5$, and $1$ in each unit cell. As
shown in Ref. \onlinecite{HongYao}, the ground state lies in the
sector of the Hilbert space where either (i) $P_{0}(m,n)=1$ and
$P_{1}(m',n')=P_{2}(m'',n'')=1$ or (ii) $P_{0}(m,n)=1$ and
$P_{1}(m',n')=P_{2}(m'',n'')=-1$. In each of these two sectors, the
time reversal symmetry of Hamiltonian (\ref{SspinH}) is broken. The
former is called as the uniform-flux sector. This means that the
ground state has another 2-fold degeneracy due to this broken time
reversal symmetry, in addition to the topological degeneracy
discussed above. Thus, we have 8-fold (6-fold) degeneracy for the
ground state in the Abelian (non-Abelian) phase, corresponding to 8
(6) flux configurations for $\Phi_{x,m},\Phi_{y,n},P_0, P_1$, and
$P_2$.

Here we focus on the ground state in the uniform-flux sector with
$(\Phi_{x,m},\Phi_{y,n})=(1,1)$, which allow both Abelian and
non-Abelian phases.\cite{HongYao} Note that only one ground state
exists in this sector. Now, Hamiltonian (\ref{HMajorana}) is reduced
to

\begin{eqnarray}
H_u&=&iJ_x\left[\sum_{x\text{-link}(\bigtriangleup)}c_{m,n}^{(3)}c_{m,n+1}^{(2)}+\sum_{x\text{-link}(\bigtriangledown)}c_{m,n}^{(5)}c_{m,n+1}^{(6)}\right]\nonumber\\
\!\!&\!\!\!\!&\!\!-iJ_y\left[\sum_{y\text{-link}(\bigtriangleup)}c_{m,n}^{(1)}c_{m,n+1}^{(3)}+\sum_{y\text{-link}(\bigtriangledown)}c_{m,n}^{(6)}c_{m,n+1}^{(4)}\right]
\nonumber\\\!\!&\!\!\!\!&\!\!+iJ_z\left[\sum_{z\text{-link}(\bigtriangleup)}c_{m,n}^{(1)}c_{m,n+2}^{(2)}+\sum_{z\text{-link}(\bigtriangledown)}c_{m,n+2}^{(4)}c_{m,n}^{(5)}\right]\nonumber\\
\!\!&\!\!\!\!&\!\!+
iJ_x'\sum_{x'\text{-link}}c_{m,n}^{(4)}c_{m,n+1}^{(1)}
-iJ_y'\sum_{y'\text{-link}}c_{m,n}^{(2)}c_{m,n+1}^{(5)}\nonumber\\\!\!&\!\!\!\!&\!\!
+iJ_z'\sum_{z'\text{-link}}c_{m,n}^{(3)}c_{m+1,n}^{(6)}.\label{Huniform}
\end{eqnarray}

By using the Fourier transform
\begin{eqnarray}
c_{\mathbf{r}}^{(s)}
&=&\sqrt{\frac{12}{NM}}\sum_{\mathbf{k}\in\mathrm{BZ}}e^{i\mathbf{k}\cdot
\mathbf{r}}c_{\mathbf{k}}^{(s)},\nonumber\\
c_{\mathbf{k}}^{(s)}&=&\sqrt{\frac{3}{NM}}\sum_{\mathbf{r}}e^{-i\mathbf{k}\cdot
\mathbf{r}}c_{\mathbf{r} }^{(s)},\label{Fourier}
\end{eqnarray}
which satisfies
\begin{eqnarray}
\left\{c_{\mathbf{r}}^{(s)},~c_{\mathbf{r'}}^{(s')}\right\}&=&2\delta_{\mathbf{r},\mathbf{r'}}\delta_{s,s'},~~
\left\{c_{\mathbf{k}}^{(s)},c_{\mathbf{k'}}^{(s')}\right\}=\delta_{\mathbf{k},-\mathbf{k'}}\delta_{s,s'},\nonumber\\
\end{eqnarray}
where $\mathrm{BZ}$ denotes the first Brillouin zone,
$s=1,~2,\cdots,~6$ denote the six sites in each unit cell, and
$\mathbf{r}$ is the position of a unit cell, Eq.~(\ref{Huniform})
becomes
\begin{eqnarray}
H_u&=&2i\sum_{\mathbf{k}\in\mathrm{BZ}}\Big\{ J_x' e^{
-i\mathbf{k}\cdot \mathbf{e_2}}
c_{\mathbf{k}}^{(4)}c_{-\mathbf{k}}^{(1)}  +J_x\left[
c_{\mathbf{k}}^{(5)}c_{-\mathbf{k}}^{(6)}+
c_{\mathbf{k}}^{(3)}c_{-\mathbf{k}}^{(2)}\right] \nonumber\\
\!\!&\!\!\!\!&\!\! -J_y' e^{ -i\mathbf{k}\cdot \mathbf{e_1}}
c_{\mathbf{k}}^{(2)}c_{-\mathbf{k}}^{(5)}-J_y\left[
c_{\mathbf{k}}^{(1)}c_{-\mathbf{k}}^{(3)}+
c_{\mathbf{k}}^{(6)}c_{-\mathbf{k}}^{(4)}\right]\nonumber\\
&& +J_z'c_{\mathbf{k}}^{(3)}c_{-\mathbf{k}}^{(6)} +J_z\left[
c_{\mathbf{k}}^{(1)}c_{-\mathbf{k}}^{(2)}+
c_{\mathbf{k}}^{(4)}c_{-\mathbf{k}}^{(5)}\right]\Big\},\label{vortexfreeH}
\end{eqnarray}
where $c_{-\mathbf{k}}^{(s)}= c_{\mathbf{k}}^{(s)\dag} $, and the
two basis vectors of the unit cell are $ \mathbf{e}_1=
\mathbf{e}_x/2- \sqrt 3 \mathbf{e}_y/2$, and
$\mathbf{e}_2=\mathbf{e}_x/2+ \sqrt 3 \mathbf{e}_y/2$. To obtain the
quasi-particle spectrum, we write Eq.~(\ref{vortexfreeH}) as
\begin{eqnarray}
H_u\!\!&\!\!=\!\!&\!\!\sum_{\mathbf{k}\in\mathrm{BZ}}\Phi_{\mathbf{k}}^{\dag}H_{\mathbf{k}}\Phi_{\mathbf{k}},\label{eq11}
\end{eqnarray}
where
$\Phi_{\mathbf{k}}^{\dag}=\left(c_{\mathbf{k}}^{(1)},c_{\mathbf{k}}^{(2)},c_{\mathbf{k}}^{(3)},
c_{\mathbf{k}}^{(4)},c_{\mathbf{k}}^{(5)},c_{\mathbf{k}}^{(6)}\right)$,
and
\begin{widetext}
\begin{eqnarray}
H_{\mathbf{k}}&=&
 \left(\begin{array}{cccccc}
0&iJ_z&-iJ_y&-iJ_x'e^{i\mathbf{k}\cdot\mathbf{e_2}}&0&0\\
 -iJ_z&0&-iJ_x&0&-iJ_y' e^{ -i \mathbf{k}\cdot\mathbf{e_1}}&0\\
  iJ_y&iJ_x&0&0&0&iJ_z'\\
 iJ_x'e^{ -i \mathbf{k}\cdot\mathbf{e_2}}&0&0&0&iJ_z&iJ_y\\
  0&iJ_y' e^{i \mathbf{k}\cdot\mathbf{e_1}}&0&-iJ_z&0&iJ_x\\
 0&0&-iJ_z'&-iJ_y&-iJ_x&0
\end{array}
 \right).\label{matrixformofH}
\end{eqnarray}
\end{widetext}
From the eigenvalue equation
$H_{\mathbf{k}}\Psi=\varepsilon_{\mathbf{k}}\Psi$, one has
\begin{eqnarray}
\varepsilon_{\mathbf{k}}^6-a\varepsilon_{\mathbf{k}}^4+b\varepsilon_{\mathbf{k}}^2-c=0,\label{eigenvalueequation}
\end{eqnarray}
where
\begin{eqnarray}
a\! & \!=\! & \!2(J_x^2+J_y^2+J_z^2)+J_x'^2+J_y'^2+J_z'^2,\nonumber\\
b\! & \!=\! &
\!2(J_x^2J_y^2+J_y^2J_z^2+J_z^2J_x^2+J_x^2J_x'^2+J_y^2J_y'^2+J_z^2J_z'^2)\nonumber\\&&+J_x'^2J_y'^2+J_y'^2J_z'^2+J_z'^2J_x'^2
+J_x^4+J_y^4+J_z^4\nonumber\\&& -2J_x'J_y'J_z^2\cos
k_x-2J_z'J_x'J_y^2\cos k_2\nonumber\\&&
-2J_y'J_z'J_x^2\cos k_1,\nonumber\\
c\! & \!=\! & \!J_x^4J_x'^2+J_y^4J_y'^2+J_z^4J_z'^2+J_x'^2J_y'^2J_z'^2\nonumber\\
&&-2J_x'J_y'\left(J_z^2J_z'^2-J_x^2J_y^2\right)\cos k_x\nonumber\\
&&-2J_y'J_z'\left(J_x^2J_x'^2-J_y^2J_z^2\right)\cos  k_1\nonumber\\
&&-2J_z'J_x'\left(J_y^2J_y'^2-J_z^2J_x^2\right)\cos k_2,
\end{eqnarray}
with
\begin{eqnarray}
k_1&=&\frac{k_x-\sqrt3k_y}{2},~k_2=\frac{k_x+\sqrt3k_y}{2}.
\end{eqnarray}
The solution of Eq.~(\ref{eigenvalueequation}) reads
\begin{eqnarray}
\varepsilon_{\mathbf{k}}^{(1)}&=&-\varepsilon_{\mathbf{k}}^{(6)}=-\sqrt{\frac{a}{3}+2p\cos\varphi},\nonumber\\
\varepsilon_{\mathbf{k}}^{(2)}&=&-\varepsilon_{\mathbf{k}}^{(5)}=-\sqrt{\frac{a}{3}
-p\left(\cos\varphi-\sqrt3\sin\varphi\right)},\nonumber\\
\varepsilon_{\mathbf{k}}^{(3)}&=&-\varepsilon_{\mathbf{k}}^{(4)}=-\sqrt{\frac{a}{3}-p\left(\cos\varphi+\sqrt3\sin\varphi\right)},\nonumber\\\label{eigenenergyAppendix}
\end{eqnarray}
where
\begin{eqnarray}
\varphi&=&\frac{1}{3}\arccos\left(\frac{q}{2p^3}\right),~~
0\leq\varphi\leq\frac{\pi}{3},\nonumber\\p&=&\frac{1}{3}\sqrt{a^2-3b}
,~~ q=\frac{2a^3}{27}-\frac{ab}{3}+c.
\end{eqnarray}

With the six energy bands in Eq.~(\ref{eigenenergyAppendix}), the
Hamiltonian (\ref{matrixformofH}) can be written as
\begin{eqnarray}
H_u& =& \sum_{j=1,\mathbf{k}\in
\mathrm{BZ}}^{j=3}\varepsilon_{\mathbf{k}}^{(j)}\left[2A_{\mathbf{k}}^{(j)\dag}A_{\mathbf{k}}^{(j)}-1\right],\label{SHfermion}
\end{eqnarray}
where
$A_{\mathbf{k}}^{(j)\dag}=\sum_{s=1}^{6}c_{\mathbf{k}}^{(s)}w_{s}(\varepsilon_{\mathbf{k}}^{(j)})
$ is a fermionic operator, with $c_{\mathbf{k}}^{(s)}$ the Fourier
transform of the Majorana fermionic operator at site $s$
($s=1,2,\cdots,6$) of the unit cell, and
\begin{eqnarray}
w_s(\varepsilon_{\mathbf{k}}^{(j)})&=&\frac{w'_s(\varepsilon_{\mathbf{k}}^{(j)})}{\sqrt{\sum_{i=1}^6|w'_i(\varepsilon_{\mathbf{k}}^{(j)})|^2}}.\label{wdy}
\end{eqnarray}
In Eq.~(\ref{wdy}),
\begin{eqnarray}
w'_1\!\!&\!\!=\!\!&\!\!
\frac{iJ_x'}{W}\left[(\varepsilon_{\mathbf{k}}^{(j)})^2-J_x^2\right]e^{ik_2}
w'_4
  +\frac{iJ_y'}{W} \left(iJ_z\varepsilon_{\mathbf{k}}^{(j)} \right.\nonumber\\&&
+J_xJ_y\Big) e^{
  -ik_1}w'_5+\frac{iJ_z'}{W}\left( iJ_y
\varepsilon_{\mathbf{k}}^{(j)}-J_xJ_z  \right) w'_6, \nonumber\\
w'_2&=&\frac{iJ_x'}{W}\left( J_xJ_y
-iJ_z\varepsilon_{\mathbf{k}}^{(j)} \right)
e^{ik_2}w'_4+\frac{iJ_y'}{W}\left[
(\varepsilon_{\mathbf{k}}^{(j)})^2\right.\nonumber\\&&-J_y^2
\Big]e^{ -ik_1}w'_5 +\frac{iJ_z'}{W} \left(J_yJ_z+iJ_x
\varepsilon_{\mathbf{k}}^{(j)}\right)
w'_6 ,\nonumber\\
w'_3&=& \frac{iJ_x'}{W}\left(J_xJ_z+iJ_y
\varepsilon_{\mathbf{k}}^{(j)} \right)
e^{ik_2}w'_4+\frac{iJ_y'}{W}\left(iJ_x
\varepsilon_{\mathbf{k}}^{(j)}\right.\nonumber\\&&-J_y J_z  \Big)
e^{ -ik_1}w'_5+\frac{iJ_z'}{W}\left[J_z^2
-(\varepsilon_{\mathbf{k}}^{(j)})^2\right]w'_6,\nonumber\\ w'_4&=&
\frac{iJ_zW-J_x'J_y'\left(iJ_z\varepsilon_{\mathbf{k}}^{(j)} +J_xJ_y
\right)e^{ -ik_x}}
{\varepsilon_{\mathbf{k}}^{(j)}W+\left[(\varepsilon_{\mathbf{k}}^{(j)})^2-J_x^2\right]J_x'^2
}  w'_5 \nonumber\\
&&+  \frac{iJ_yW-J_z'J_x'\left(iJ_y
\varepsilon_{\mathbf{k}}^{(j)}-J_xJ_z \right)e^{ -ik_2}}
{\varepsilon_{\mathbf{k}}^{(j)}W
+\left[(\varepsilon_{\mathbf{k}}^{(j)})^2-J_x^2\right]J_x'^2 }
w'_6,\nonumber\\
w'_5  &=&iJ_xJ_x'^2J_y'J_z' e^{ ik_1}-J_y'J_z' \left(J_yJ_z+iJ_x
\varepsilon_{\mathbf{k}}^{(j)}\right)\varepsilon_{\mathbf{k}}^{(j)}
e^{ik_1}\nonumber\\&& +iJ_x \left\{\varepsilon_{\mathbf{k}}^{(j)}W
+\left[(\varepsilon_{\mathbf{k}}^{(j)})^2-J_x^2\right]J_x'^2
\right\}\nonumber\\&& +J_yJ_zW + iJ_zJ_z'J_x'\left( -J_xJ_z +iJ_y
\varepsilon_{\mathbf{k}}^{(j)} \right)e^{
-ik_2}\nonumber\\&&-iJ_yJ_x'J_y'\left(J_xJ_y
-iJ_z\varepsilon_{\mathbf{k}}^{(j)}\right) e^{i
k_x},\nonumber\\
 w'_6 &=&
\left[(J_x'^2+J_y'^2)(\varepsilon_{\mathbf{k}}^{(j)})^2-J_x'^2J_y'^2-J_x^2J_x'^2-J_y^2J_y'^2\right]
\varepsilon_{\mathbf{k}}^{(j)}\nonumber\\&&
+\left[(\varepsilon_{\mathbf{k}}^{(j)})^2-J_z^2\right]W -J_zJ_x'J_y'
\Big( 2J_xJ_y\sin{ k_x}\nonumber\\&&
-2J_z\varepsilon_{\mathbf{k}}^{(j)}\cos{k_x}\Big), \label{w1to6}
\end{eqnarray}
with
\begin{eqnarray}
W&=&
\left[J_x^2+J_y^2+J_z^2-(\varepsilon_{\mathbf{k}}^{(j)})^2\right]\varepsilon_{\mathbf{k}}^{(j)}.
\end{eqnarray}

In order to find the ground-state wavefunction for Hamiltonian
(\ref{SHfermion}), we should first define the vacuum $|0\rangle$ by
\begin{equation}
a_{m,n}|0\rangle=0,
\end{equation}
where $a_{m,n}$ is the fermionic operator defined in
Eq.~(\ref{Jordan}), and the index $(m,n)$ runs over every site of
the lattice. From Eqs.~(\ref{Majorana1}) and (\ref{Majorana2}), we
have
\begin{eqnarray}
\left\langle 0\left|c_{\mathbf{r}}^{(s)} c_{\mathbf{r}'}^{(s')}
\right|0\right\rangle=\delta_{\mathbf{r}',\mathbf{r}}\delta_{s',s}.
\end{eqnarray}
Through Fourier transform, it leads to
\begin{eqnarray}
\left\langle 0\left|c_{\mathbf{k}}^{(s)} c_{\mathbf{k}'}^{(s')}
\right|0\right\rangle
&=&\frac{1}{2}\delta_{\mathbf{k}',-\mathbf{k}}\delta_{s',s}.\label{oneover2}
\end{eqnarray}
For state
$|j,\mathbf{k}\rangle=\sqrt{2}A_{\mathbf{k}}^{(j)\dag}|0\rangle$ in
the occupied band, it is normalized as
\begin{eqnarray}
\langle j,\mathbf{k}|j,\mathbf{k}\rangle
&=&\langle0|\sqrt{2}A_{\mathbf{k}}^{(j)} \sqrt{2}A_{\mathbf{k}}^{(j)\dag}|0\rangle\nonumber\\
&=&2\sum_{s,s'=1}^{6}w_{s}^{\ast}(\varepsilon_{\mathbf{k}}^{(j)})
w_{s'}(\varepsilon_{\mathbf{k}}^{(j)})\left\langle0\left|c_{-\mathbf{k}}^{(s)}c_{\mathbf{k}}^{(s')}\right|0\right\rangle\nonumber\\
&=&2\sum_{s,s'=1}^{6}w_{s}^{\ast}(\varepsilon_{\mathbf{k}}^{(j)})
w_{s'}(\varepsilon_{\mathbf{k}}^{(j)})\cdot\frac{1}{2}\delta_{s',s}\nonumber\\
&=&\sum_{s=1}^{6}|w_{s}(\varepsilon_{\mathbf{k}}^{(j)})|^2\nonumber\\
&=&1,
\end{eqnarray}
which clarifies that the normalization factor $\sqrt2$ is due to the
fact that the occupation number can only take $1/2$ for the Fourier
transform of each Majorana fermion [see Eq. (\ref{oneover2})]. Hence
we can write the ground-state wavefunction as
\begin{eqnarray}
|g\rangle=\prod_{\mathbf{k}\in\mathrm{BZ}}\prod_{j=1}^3\sqrt{2}A_{\mathbf{k}}^{(j)\dag}|0\rangle.
\end{eqnarray}
The corresponding ground-state energy per site is
\begin{eqnarray}
 E_g
&=&\frac{1}{6MN}\sum_{\mathbf{k}\in\mathrm{BZ}}\left(\varepsilon_{\mathbf{k}}^{(1)}+\varepsilon_{\mathbf{k}}^{(2)}+\varepsilon_{\mathbf{k}}^{(3)}\right)\nonumber\\
&=&\frac{\sqrt3}{48\pi^2}\int_{\mathrm{BZ}}
d^2k\left(\varepsilon_{\mathbf{k}}^{(1)}+\varepsilon_{\mathbf{k}}^{(2)}+\varepsilon_{\mathbf{k}}^{(3)}\right).
\end{eqnarray}

\subsection{Two different ground states in the same topological
class}

Below we show that the quantum phase transition (QPT) between two
phases belonging to the same topological class is nontrivial. For
example, in the case of $J_x=J_y,~J_x'=J_y'$, and $J_zJ_x'=J_z'J_x$,
the QPT that occurs when $\Lambda_1\equiv J_x'/J_x$ varies across
the critical line $\Lambda_1=0$ does not change the topological
class of the ground state, i.e., the QPT occurs between either two
non-Abelian phases with Chern number $\nu=\pm1$ or two Abelian
phases with $\nu=0$. By using the parameters $\Lambda_1\equiv
J_x'/J_x$ and $\Lambda_2\equiv J_z/J_x$, Eq.~(\ref{w1to6}) can be
reduced to
\begin{eqnarray}
 w'_1\!&\!=\!&\!  \frac{i\Lambda_1}
{W'}
\left\{\left[(\varepsilon_{\mathbf{k}}^{(j)})^2-1\right]e^{ik_2}
w'_4+ \left(i\Lambda_2\varepsilon_{\mathbf{k}}^{(j)} +1\right)e^{
-ik_1}w'_5\right.\nonumber\\&&+\left.\Lambda_2\left(i
\varepsilon_{\mathbf{k}}^{(j)}-\Lambda_2 \right)w'_6\right\}, \nonumber\\
w'_2 \!&\!=\!&\!\frac{i\Lambda_1  } {W' }  \left\{\left(1
-i\Lambda_2\varepsilon_{\mathbf{k}}^{(j)}\right) e^{ik_2} w'_4
+\left[(\varepsilon_{\mathbf{k}}^{(j)})^2-1\right]e^{ -ik_1} w'_5
\right.\nonumber\\&&\left.+\Lambda_2\left(\Lambda_2+i
\varepsilon_{\mathbf{k}}^{(j)} \right)
 w'_6 \right\},\nonumber\\  w'_3 \!&\!=\!&\! \frac{
i\Lambda_1 } {W' }  \left\{\left(\Lambda_2+i
\varepsilon_{\mathbf{k}}^{(j)}\right)e^{ik_2} w'_4 +\left(i
\varepsilon_{\mathbf{k}}^{(j)}-\Lambda_2\right) e^{ -ik_1}\right.
w'_5\nonumber\\&& \left.+\Lambda_2\left[\Lambda_2^2
-(\varepsilon_{\mathbf{k}}^{(j)})^2\right] w'_6 \right\},\nonumber\\
 w'_4 \!&\!=\!&\!\frac{i\Lambda_2W'-\Lambda_1^2\left(i\Lambda_2\varepsilon_{\mathbf{k}}^{(j)}
+1 \right)e^{ -ik_x} } {W'
\varepsilon_{\mathbf{k}}^{(j)}+\left[(\varepsilon_{\mathbf{k}}^{(j)})^2-1\right]\Lambda_1^2
} w'_5  \nonumber\\
&&+ \frac{i W'-\Lambda_2\left(i \varepsilon_{\mathbf{k}}^{(j)}
-\Lambda_2\right)e^{ -ik_2} } {W'
\varepsilon_{\mathbf{k}}^{(j)}+\left[(\varepsilon_{\mathbf{k}}^{(j)})^2-1\right]\Lambda_1^2
}  w'_6 ,\nonumber\\ w'_5 \!&\!=\!&\!
J_x^5\Big\{i\Lambda_1^4\Lambda_2 e^{ ik_1}-\Lambda_1^2\Lambda_2
\left(\Lambda_2+i \varepsilon_{\mathbf{k}}^{(j)}\right)
\varepsilon_{\mathbf{k}}^{(j)} e^{ik_1} +\Lambda_2W'\nonumber\\&&
+i\Lambda_1^2\Lambda_2^2\left( i
\varepsilon_{\mathbf{k}}^{(j)}-\Lambda_2 \right)e^{ -ik_2}
-i\Lambda_1^2\left(1
-i\Lambda_2\varepsilon_{\mathbf{k}}^{(j)}\right)
e^{ik_x}\nonumber\\&&+i \left[W'
\varepsilon_{\mathbf{k}}^{(j)}+\Lambda_1^2
(\varepsilon_{\mathbf{k}}^{(j)})^2-\Lambda_1^2 \right]\Big\},
\nonumber\\  w'_6 \!&\!=\!&\!J_x^5\Big\{
\Lambda_1^2\left[2(\varepsilon_{\mathbf{k}}^{(j)})^2-
\Lambda_1^2-2\Lambda_2^2\right] \varepsilon_{\mathbf{k}}^{(j)}
+\left[(\varepsilon_{\mathbf{k}}^{(j)})^2\right.\nonumber\\
&&-\Lambda_2^2\Big]W'  -\Lambda_1^2\Lambda_2 \left(
2\sin{k_x}-2\Lambda_2\varepsilon_{\mathbf{k}}^{(j)}\cos{k_x}\right)\Big\},
\end{eqnarray}
with
\begin{eqnarray}
W'&=&\left[2+\Lambda_2^2-(\varepsilon_{\mathbf{k}}^{(j)})^2\right]\varepsilon_{\mathbf{k}}^{(j)},
\end{eqnarray}
where $\varepsilon_{\mathbf{k}}^{(j)}$ is in units of $J_x$. These
$w'_s$ satisfy
\begin{eqnarray}
w'_s(-\Lambda_1,\Lambda_2)&=&-w'_s(\Lambda_1,\Lambda_2),~s=1,2,3;\nonumber\\
w'_s(-\Lambda_1,\Lambda_2)&=&w'_s(\Lambda_1,\Lambda_2),~s=4,5,6.\label{wminusplus}
\end{eqnarray}
From $
A_{\mathbf{k}}^{(i)\dag}=\sum_{s=1}^{6}c_{\mathbf{k}}^{(s)}w_{s}(\varepsilon_{\mathbf{k}}^{(j)})
$, where $w_{s}$ are given in Eq.~(\ref{wdy}), we can define an
analytical function
$F\left(c_{\mathbf{k}}^{(1)},c_{\mathbf{k}}^{(2)}
,c_{\mathbf{k}}^{(3)} ,c_{\mathbf{k}}^{(4)} , c_{\mathbf{k}}^{(5)}
,c_{\mathbf{k}}^{(6)} \right)$ by
\begin{eqnarray}
|g(\Lambda_1,\Lambda_2)\rangle&=&\prod_{\mathbf{k}\in
\mathrm{BZ}}\prod_{j=1}^3\sqrt{2}A_{\mathbf{k}}^{(j)\dag}|0\rangle\nonumber\\
&\equiv&\prod_{\mathbf{k}\in
\mathrm{BZ}}F\left(c_{\mathbf{k}}^{(1)},c_{\mathbf{k}}^{(2)}
,c_{\mathbf{k}}^{(3)} ,c_{\mathbf{k}}^{(4)} , c_{\mathbf{k}}^{(5)}
,c_{\mathbf{k}}^{(6)}
\right)\Big|0\Big\rangle.\nonumber\\&~&\label{39}
\end{eqnarray}
From Eq.~(\ref{wminusplus}), it follows that
\begin{eqnarray}
&&|g(-\Lambda_1,\Lambda_2)\rangle \nonumber\\
&&=\prod_{\mathbf{k}\in
\mathrm{BZ}}F\left(-c_{\mathbf{k}}^{(1)},-c_{\mathbf{k}}^{(2)}
,-c_{\mathbf{k}}^{(3)} ,c_{\mathbf{k}}^{(4)} , c_{\mathbf{k}}^{(5)}
,c_{\mathbf{k}}^{(6)} \right)\Big|0\Big\rangle.~~~~~~ \label{40}
\end{eqnarray}
Equations (\ref{39}) and (\ref{40}) show that the ground state $
|g(\Lambda_1,\Lambda_2)\rangle$ is mapped to $
|g(-\Lambda_1,\Lambda_2)\rangle$ when $c_{\mathbf{k}}^{(s)}$,
$s=1,2$, and $3$, are changed to $-c_{\mathbf{k}}^{(s)}$. Because
$c_{\mathbf{k}}^{(s)},~s=1,2$, and $3$, are the Fourier transform of
the Majorana fermions $c_{\mathbf{r}}^{(s)}$ defined on sites $1,2$,
and $3$ in each unit cell (i.e., all the triangles related to
plaquette operator $P^{(2)}$). The transformation
$c_{\mathbf{k}}^{(s)}\rightarrow -c_{\mathbf{k}}^{(s)}$ corresponds
to a $\pi$ phase change to each of the Majorana fermions
$c_{\mathbf{r}}^{(s)}$ at sites $s=1,2$, and $3$. Note that the two
ground states $ |g(\Lambda_1,\Lambda_2)\rangle$ and $
|g(-\Lambda_1,\Lambda_2)\rangle$ are different [see Eqs.~~(\ref{39})
and (\ref{40})], although the energy spectrum of the Hamiltonian
$H(\Lambda_1,\Lambda_2)$ is symmetric about the critical line
$\Lambda_1=0$.

\section{Effective Hamiltonian}

In order to show that both the QPTs between two Abelian phases and
those between two non-Abelian phases are nontrivial, we use the
perturbation method in Ref. \onlinecite{Kitaev06} to derive the
effective Hamiltonian at $\Lambda_1\sim0$. In this case with
$J_x=J_y,~J_x'=J_y'$, and $J_zJ_x'=J_z'J_x$, we characterize the
phase diagram by the two parameters $\Lambda_1\equiv J_x'/J_x$, and
$\Lambda_2\equiv J_z/J_x$. We assume $J_x>0$ and take it as the unit
of energy.

The Hamiltonian of the system can be written as
\begin{eqnarray}
H&=&H_0+V,\label{s301}
\end{eqnarray}
with
\begin{eqnarray}
H_0&=&\sum_{x\text{-link}}\sigma_i^x\sigma_j^x
+\sum_{y\text{-link}}\sigma_i^y\sigma_j^y+\Lambda_2\sum_{z\text{-link}}\sigma_i^z\sigma_j^z,
\nonumber\\
V&=&\Lambda_1\sum_{x'\text{-link}}\sigma_i^x\sigma_j^x
+\Lambda_1\sum_{y'\text{-link}}\sigma_i^y\sigma_j^y+\Lambda_1\Lambda_2\sum_{z'\text{-link}}\sigma_i^z\sigma_j^z,\nonumber\\
\end{eqnarray}
where $V$ is a perturbation.

As in Ref. \onlinecite{Kitaev06}, we calculate the effective
Hamiltonian as
\begin{eqnarray}
H_{\mathrm{eff}}&=&\Upsilon^{\dag}(V+VG_0'V+VG_0'VG_0'V+\cdots)\Upsilon\nonumber\\
&=&H^{(1)}+H^{(2)}+H^{(3)}+\cdots,\label{s302}
\end{eqnarray}
where $\Upsilon$ maps the effective Hilbert space onto the
ground-state subspace of $H_0$ and $G_0'$ is the Green function for
excited states of $H_0$. The ground state of $H_0$ is a
direct-product state consisting of the ground states of the
following Hamiltonians:
\begin{eqnarray}
H_{p_1}&=&\sigma_1^x\sigma_3^x +\sigma_2^y\sigma_3^y
+\Lambda_2\sigma_1^z\sigma_2^z,
\nonumber\\
H_{p_2}&=&\sigma_2^x\sigma_3^x +\sigma_1^y\sigma_2^y
+\Lambda_2\sigma_1^z\sigma_3^z,
\end{eqnarray}
defined on down-pointing and up-pointing triangles [see
Fig.~\ref{Fisher}(b)].

In Eq.~(\ref{s302}), $H^{(1)}=\Upsilon^{\dag}V\Upsilon$ is obtained
as
\begin{eqnarray}
H ^{(1)}&=&\Lambda_1\big(\sum_{x'\text{-link}}\sigma_i^x\sigma_j^x
+\sum_{y'\text{-link}}\sigma_i^y\sigma_j^y+\Lambda_2\sum_{z'\text{-link}}\sigma_i^z\sigma_j^z\big).\nonumber\\
\end{eqnarray}
As in the perturbation method used for the Kitaev model on a
honeycomb lattice in the presence of a week magnetic
field,\cite{Kitaev06} in order to derive
$H^{(2)}=\Upsilon^{\dag}VG_0'V\Upsilon$, we assume that all the
involved excited states of the system have a gap
$\varepsilon\sim|\Lambda_2|$ above the ground-state energy. This
gives rise to
\begin{eqnarray}
H
^{(2)}&=&-\frac{MN\Lambda_1^2}{6|\Lambda_2|}(2+\Lambda_2^2)\nonumber\\
&&-\frac{\Lambda_1^2}{|\Lambda_2|}\big(\sum_{x'\text{-link}}\sigma_i^x\sigma_j^x
\big)\big( \sum_{y'\text{-link}}\sigma_k^y\sigma_l^y\big)\nonumber\\
&&-\frac{\Lambda_1^2\Lambda_2}{|\Lambda_2|}\big(\sum_{x'\text{-link}}\sigma_i^x\sigma_j^x
\big)\big( \sum_{z'\text{-link}}\sigma_k^z\sigma_l^z\big)\nonumber\\
&&-\frac{\Lambda_1^2\Lambda_2}{|\Lambda_2|}\big(\sum_{y'\text{-link}}\sigma_i^y\sigma_j^y
\big)\big( \sum_{z'\text{-link}}\sigma_k^z\sigma_l^z\big).\label{B6}
\end{eqnarray}
It is interesting to note that though $H ^{(2)}$ involves even
number of spin operators, it is not invariant under a time reversal
transformation in the uniform-flux sector. Take one sum in the
second line of Eq.~(\ref{B6}) as an example:
\begin{eqnarray}
H ^{(2)}_p&=&-\frac{\Lambda_1^2}{|\Lambda_2|}\sum_{\langle
jk\rangle=
z\text{-link}}\sigma_i^x\sigma_j^x\sigma_k^y\sigma_l^y\nonumber\\
&=&-\frac{\Lambda_1^2}{|\Lambda_2|}\sum_{\langle jk\rangle=
z\text{-link}}(\sigma_j^x\sigma_k^y\sigma_m^z)\sigma_m^z\sigma_i^x\sigma_l^y,\label{B12}
\end{eqnarray}
where $(j,k,m)$ denotes the three sites in either a down-pointing or
up-pointing triangle, which is labeled by either $(2,1,3)$ or
$(1,3,2)$ in Fig.~\ref{Fisher}(b). Namely,
$\sigma_j^x\sigma_k^y\sigma_m^z$ is either the plaquette operator
$P_1$ or $P_2$. Since the eigenvalues of $P_1$ and $P_2$ are both
chosen to be $1$ in the uniform-flux sector, $H ^{(2)}_p$ can be
reduced to involve only three spin operators in this sector and the
time reversal symmetry is now broken. Thus, in the uniform-flux
sector, one has
\begin{eqnarray}
\mathcal{T}H ^{(2)}\mathcal{T}^{-1}\neq H ^{(2)},\label{B15}
\end{eqnarray}
where $\mathcal{T}$ is the time reversal transformation. Similarly,
$H^{(3)}=\Upsilon^{\dag}VG_0'VG_0'V\Upsilon$ can be obtained as
\begin{eqnarray}
H ^{(3)}&=&\frac{\Lambda_1^3(1+\Lambda_2^2)MN}{3\Lambda_2^2}
\big(\sum_{x'\text{-link}}\sigma_i^x\sigma_j^x
+\sum_{y'\text{-link}}\sigma_i^y\sigma_j^y\big)\nonumber\\
&&+\frac{2\Lambda_1^3MN}{3\Lambda_2}
\sum_{z'\text{-link}}\sigma_i^z\sigma_j^z\nonumber\\
&&+\frac{6\Lambda_1^3}{\Lambda_2}
\sum_{(i_1,i_2,i_3)\neq\bigtriangleup ,\bigtriangledown}
\sigma_{i_1}^x\sigma_{j_1}^x\sigma_{i_2}^y\sigma_{j_2}^y\sigma_{i_3}^z\sigma_{j_3}^z\nonumber\\
&&+H_p^{(3)}.
\end{eqnarray}
Here $(i_1,j_1)$ are the two sites connected by an $x'$-link,
$(i_2,j_2)$ are the two sites connected by a $y'$-link, and
$(i_3,j_3)$ are the two sites connected by a $z'$-link.
\begin{eqnarray}
H_p ^{(3)}&=& \frac{6\Lambda_1^3}{\Lambda_2} \sum_{
\bigtriangledown}P_1(m,n)\sigma_{m,n-1}^y\sigma_{m,n+3}^x\sigma_{m-1,n+1}^z\nonumber\\
&&+\frac{6\Lambda_1^3}{\Lambda_2} \sum_{
\bigtriangleup}P_2(m,n)\sigma_{m,n-1}^x\sigma_{m,n+3}^y\sigma_{m+1,n+1}^z,\nonumber\\
\end{eqnarray}
where $\bigtriangledown (\bigtriangleup)$ runs over all
down-pointing (up-pointing) triangles with $(m,n)$ denoting the site
$5$ ($1$) in each unit cell (see Fig.~\ref{Fisher2}). By summing
over $H^{(1)},H^{(2)}$ and $H^{(3)}$, the effective Hamiltonian, up
to the third order, is given by
\begin{eqnarray}
H_{\mathrm{eff}}&=&H_{0}^{(3)}+\frac{6\Lambda_1^3}{\Lambda_2} \sum_{
\bigtriangledown}P_1(m,n)\sigma_{m,n-1}^y\sigma_{m,n+3}^x\sigma_{m-1,n+1}^z\nonumber\\
&&+\frac{6\Lambda_1^3}{\Lambda_2} \sum_{
\bigtriangleup}P_2(m,n)\sigma_{m,n-1}^x\sigma_{m,n+3}^y\sigma_{m+1,n+1}^z,\nonumber\\
\end{eqnarray}
where $H_{0}^{(3)}$ is a term involving none of the plaquette
operators $P_i$, $i=0,1$ and $2$.

For the QPT between two phases of the same Chern number, the
parameter $\Lambda_1$ changes its sign at the transition point. This
yeilds the change of signs in certain terms in the effective
Hamiltonian. Therefore, the effective Hamiltonian $H_{\mathrm{eff}}$
is different at the two sides of the transition point $\Lambda_1=0$.


\begin{thebibliography}{}
\bibitem{footnote1} It is shown in Ref. \onlinecite{HongYao} that when both $J_{\alpha}$ and $J_{\alpha}'$ are positive, the
ground state is either in the uniform-flux sector or in the sector
related by the time reversal transformation. Based on this, one can
easily prove that this applies to the cases with arbitrary
$J_{\alpha}$ and $J_{\alpha}'$.
\end{thebibliography}
\end{document}